# Energy dependence of source geometry and chaoticity in hadronic collisions from Bose-Einstein correlations

I.V. Andreev[1,2*], M. Plümer[1†], B.R. Schlei[1‡] and R.M. Weiner[1§]

[1] Physics Department, Univ. of Marburg, Marburg, FRG

[2] P.N. Lebedev Institute, Moscow, Russia

## Abstract

We compare and analyse Bose-Einstein correlation data for $\pi^+ p$ and $K^+ p$ collisions at $\sqrt{s} = 22\ GeV$ obtained by the NA22-Collaboration and data for $p\bar{p}$ collisions at $\sqrt{s} = 630\ GeV$ obtained by the UA1-Minimum-Bias-Collaboration. Using a parametrization for a longitudinally expanding source, we observe that correlation lengths and radii extracted from fits to the data change significantly if one goes from the NA22- to the UA1-data: the transverse radius of the chaotic source increases by about 50% while the correlation lengths in longitudinal and transverse directions decrease by about 50% and 30%, respectively. The chaoticity parameter remains approximately constant or increases.

---

[*] E. Mail: andreev@lpi.ac.ru

[†] E. Mail: pluemer_m@vax.HRZ.Uni-Marburg.DE

[‡] E. Mail: schlei@convex.HRZ.Uni-Marburg.DE

[§] E. Mail: weiner@vax.HRZ.Uni-Marburg.DE

# 1   Introduction

Particle interferometry constitutes an important method to obtain information about space-time aspects of multiparticle production, as well as information about the quantum statistical nature of the source (coherence versus chaoticity), in high energy collisions. Since the majority of secondaries produced in such reactions are mesons, one usually considers Bose-Einstein correlation (BEC) data. In particular, one may study the dependence of radii, lifetimes, correlation lengths and the chaoticity of the particle source on the type of reaction and on the collision energy $\sqrt{s}$. This is important, among other things, for the determination of energy densities in the search for the quark-gluon-plasma. The results of such a comprehensive analysis of BEC data may provide useful constraints for phenomenological models of multiparticle production.

Note that in practice the extraction of space-time parameters and chaoticity from experimental data is not at all straightforward. To see why that is so, consider, for instance, the two-particle BEC function

$$C_2(\vec{k}_1, \vec{k}_2) = \frac{\rho_2(\vec{k}_1, \vec{k}_2)}{\rho_1(\vec{k}_1)\rho_1(\vec{k}_2)} \quad (1)$$

where $\rho_1(\vec{k})$ and $\rho_2(\vec{k}_1, \vec{k}_2)$ are the one- and two-particle inclusive spectra, respectively. In the most general case, $C_2(\vec{k}_1, \vec{k}_2)$ depends on the three-momenta $\vec{k}_i$ ($i = 1, 2$) of the two particles, i.e., on 6 independent variables (5 variables for a central collision). Because of limited statistics, the experimental data are often analysed and presented as a function of the single variable $Q^2 \equiv -q^\mu q_\mu$, where $q^\mu \equiv k_1^\mu - k_2^\mu$ is the difference of the four-momenta of the particles. Compared to other single variables (such as the rapidity difference or the difference of transverse momenta) $Q^2$ has two important advantages: (a) it is Lorentz invariant and (b) for $Q^2 = 0$, the two particles have equal momenta and therefore the BEC is maximal. The representation of the BEC data as a function of one variable only implies an integration over the remaining momentum components. In the variable $Q^2$, the two-particle BEC function reads

$$\tilde{C}_2(Q^2) = \frac{\int_\Omega d\omega_1 \int_\Omega d\omega_2 \; \rho_2(\vec{k}_1, \vec{k}_2) \; \delta\left[Q^2 + (k_1^\mu - k_2^\mu)^2\right]}{\int_\Omega d\omega_1 \int_\Omega d\omega_2 \; \rho_1(\vec{k}_1)\rho_1(\vec{k}_2) \; \delta\left[Q^2 + (k_1^\mu - k_2^\mu)^2\right]} \qquad (2)$$

where $d\omega \equiv d^3k/(2\pi)^3 2E$ is the invariant phase space volume element.

Thus, in order to obtain information about the space-time characteristics and chaoticity of the source from an analysis of the BEC data as a function of $Q^2$, it is necessary to explicitly perform the phase space integrations in (2) and to carefully take into account the kinematical cuts of the experiment under consideration[1]. In Refs. [1, 2], this procedure was applied to the new BEC-data in $p\bar{p}$ collisions at $\sqrt{s} = 630\ GeV$ obtained by the UA1-Collaboration [3], which cover a range in $Q^2$ down to values as low as $10^{-3}\ GeV^2$.

Recently, the NA22-Collaboration has measured the two-particle correlation in the variable $Q^2$ for $Q^2 \geq 10^{-3}\ GeV^2$ for $\pi^+ p$ and $K^+ p$ collisions at $\sqrt{s} = 22\ GeV$ [4]. Since both experiments cover the same wide range in $Q$, it becomes possible for the first time to compare BEC data for hadronic collisions at different CM energies and to extract information about the change in the space-time parameters and chaoticity of the particle source. It is the purpose of the present paper to do this. In section 2, we briefly introduce the formalism and our model for a longitudinally expanding source. The fits to the data and the corresponding differences in the source parameters obtained for the NA22- and the UA1-data are presented in section 3. Finally, we discuss our results and compare them to the results of other theoretical and experimental investigations in section 4.

---

[1]Note that these integrations may considerably change the shape of the correlation function. In Refs. [1, 2] it was demonstrated that a Gaussian form of $C_2(\vec{k}_1, \vec{k}_2)$ can lead to a correlation in $Q^2$ which looks like a power law.

## 2 The Expanding Source Model

In this section we briefly review the quantum statistical formalism and the model of a boost-invariant longitudinally expanding source[2,3] which were derived in [5, 6] and applied to the analysis of UA1-data in [1, 2].

It is assumed that the system can be described by a Gaussian density matrix[4]. In this case, all multiparticle distributions can be expressed in terms of two functions $I(k)$ and $D(k_1, k_2)$. In particular, the one- and two-particle inclusive distributions take the form

$$\rho_1(\vec{k}) = D(k,k) + |I(k)|^2 \tag{3}$$

$$\rho_2(\vec{k}_1, \vec{k}_2) = \rho_1(\vec{k}_1)\rho_1(\vec{k}_2)$$
$$+ 2 \text{ Re } D(k_1, k_2)I^\star(k_1)I(k_2) + D(k_1, k_2)D(k_2, k_1) \tag{4}$$

To describe the space-time properties of the source we shall work in the current formalism [9]. The functions $I(k)$ and $D(k_1, k_2)$ are the on-shell Fourier transforms of the coherent component of the current, $I(x)$, and the chaotic current correlator, $D(x, x')$, respectively, which can be written as [5, 6]

$$I(x) = f_c(x) \tag{5}$$

$$D(x, x') = f_{ch}(x) \, C(x - x') \, f_{ch}(x') \tag{6}$$

Here, $f_{ch}(x)$ and $f_c(x)$, are the space-time distributions of the chaotic and of the coherent component, respectively, and $C(x - x')$ is the primordial correlator which describes the correlation for an infinitely extended source.

---

[2] Such a model is based on the scenario of an inside-outside cascade and the presence of a plateau in the central rapidity region.

[3] We have also considered a model of a "static" (non-expanding) source [5, 6] and applied it to the new BEC data. It turns out that the static source model fails to describe the UA1-data. Our results obtained for the static source will be published in [7].

[4] The Gaussian form of the density matrix follows from the central limit theorem in the case of a large number of independent source elements[8].

For the model of a longitudinally expanding source [6], we introduce the variables

$$\tau = \sqrt{x_0^2 - x_\parallel^2}, \qquad \eta = \frac{1}{2} \ln \frac{x_0 + x_\parallel}{x_0 - x_\parallel} \qquad (7)$$

where $\tau$ is the longitudinal proper time and $\eta$ the space-time rapidity. For the sake of mathematical simplicity and in order to keep the number of parameters as low as possible, we consider for the quantities $f_{ch}(x)$, $f_c(x)$ and $C(x - x')$ forms which are invariant under boosts of the coordinate frame in longitudinal direction:

$$f_{ch}(x) \propto \delta(\tau - \tau_0) \exp\left(-\frac{x_\perp^2}{R_{ch,\perp}^2}\right) \qquad (8)$$

$$f_c(x) \propto \delta(\tau - \tau_0) \exp\left(-\frac{x_\perp^2}{R_{c,\perp}^2}\right) \qquad (9)$$

$$C(\eta - \eta', x_\perp - x'_\perp) \propto \exp\left[-\frac{2\tau_0^2}{L_\eta^2} \sinh^2\left(\frac{\eta - \eta'}{2}\right) - \frac{(\vec{x}_\perp - \vec{x}'_\perp)^2}{2L_\perp^2}\right] \qquad (10)$$

The parameters for the expanding source are[5] the proper formation time $\tau_0$, the transverse radii of the chaotic and the coherent source, $R_{ch,\perp}$ and $R_{c,\perp}$, the correlation lengths in longitudinal and transverse direction, $L_\eta$ and $L_\perp$, and the value $p_0$ of the chaoticity parameter at some fixed momentum (e.g., at $k_\perp = 0$, $p(k_\perp = 0) = p_0$). As was discussed in detail in [2], the number of free parameters can be reduced by relating them to the shape of the one-particle inclusive distribution $\rho_1(\vec{k})$. For the expanding source, the remaining four free parameters are[2] $\tau_0$, $R_{ch,\perp}$, $L_\eta$ and $p$ (where it has been assumed that the transverse momentum spectra of the chaotic and the coherent component have the same shape, in which case the chaoticity $p$ becomes independent of $\vec{k}$, cf. [1, 2]).

---

[5] We have made the simplifying assumptions that the chaotically and the coherently produced particles are emitted at the same proper time $\tau_0$ and that there are no fluctuations in $\tau_0$ (for the more general case, see Ref. [2]).

## 3  Fits to the Data

Fig. 1a shows the UA1-data[6] for the two-particle BEC as a function of $Q^2$ together with the fits obtained in Refs. [1, 2] for a longitudinally expanding, purely chaotic source ($p = 1$). The corresponding values of $\tau_0$, $R_{ch,\perp}$ and $L_\eta$, and the values of $\chi^2/d.o.f.$, are listed in Table 1.

The NA22-data are plotted in Fig. 1b. Clearly, the BEC function $\tilde{C}_2(Q^2)$ at $\sqrt{s} = 22\ GeV$ is broader than the one at 630 $GeV$. Naively, one might conclude from this observation that the radii and/or lifetimes of the source increase with CM energy. However, one must bear in mind that the correlation functions were obtained in different phase space regions $\Omega$ which enter into the integrations in (2): for the NA22-data, $\Omega$ is defined through the cuts $|y| < 2$, $k_\perp > 0$, whereas for the UA1-data one has $|y| < 3$, $k_\perp > 0.15\ GeV$. It turns out that, if one uses for the NA22-data the same parameter values as for the fit to the UA1-data (taking into account the different phase space cuts), one obtains unacceptable $\chi^2/d.o.f.$ values, i.e., one fails to reproduce the NA22-data. This implies that the observed difference in the widths of $\tilde{C}_2(Q^2)$ in the NA22- and in the UA1-experiment cannot be explained through the difference in the rapidity and transverse momentum cuts alone. Rather, it reflects a genuine difference in the geometry and quantum statistical properties of the particle sources. To describe the NA22-data, one needs to choose a set of parameter values different from the ones used to fit the UA1-data. The resultant space-time parameters are listed in Table 1, and the corresponding correlation function is compared to the NA22-data in Fig. 1b.

Up to this point, we have assumed that the particles are emitted from a purely chaotic source. Before we can discuss the quality of the fits and the $s$-dependence of the quantum statistical parameters of the source, we have to address the question if the data allow any conclusions concerning the presence of a coherent component and its energy dependence. It turns out that for the NA22-data fits with acceptable $\chi^2/d.o.f.$ can be obtained for a wide range of chaoticity parameters, $0.3 \leq p \leq 1$. For the UA1-data, the range is $0.8 \leq p \leq 1$. Fortunately, the values of the other fit parameters ($\tau_0$, $R_{ch,\perp}$ and $L_\eta$) do not strongly depend on the value of $p$. To illustrate this, we present the results of the best fit to the NA22-data for $p = 0.6$ in Table 1.

---

[6]We have normalized both the UA1-data [3] and the NA22-data [4] to unity at $Q^2 = 1\ GeV^2$ in order to subtract the effects of long range correlations which were shown to play an important role at UA1-energies[10].

Of course it would be interesting to compare the BEC data at $\sqrt{s} = 22\ GeV$ and at $\sqrt{s} = 630\ GeV$ *in the same momentum space region*. As was mentioned above, the present kinematical cuts are $|y| < 2$, $k_\perp > 0$ for the NA22-data and $|y| < 3$, $k_\perp > 0.15\ GeV$ for the UA1-data. We would therefore urge both experimental groups to re-analyze their data and present $\tilde{C}_2(Q^2)$ in the restricted phase space region $|y| < 2$, $k_\perp > 0.15\ GeV$. This would allow to explicitly see the $s$-dependence of the correlation function without having to take into account the effects of different phase space cuts.

Note that here we have not treated separately the contributions from the different resonance decays[7]. Thus the radii and lifetimes extracted from the BEC data and listed in Table 1 must be regarded as effective radii and lifetimes in so far as they represent an average of the direct pion component and the components related to pions produced from resonance decays. We have also not attempted to correct for the effects of particle misidentification.

## 4  Discussion of the Results

We have analysed the new BEC data obtained for $\pi^+ p$ and $K^+ p$ collisions at $\sqrt{s} = 22\ GeV$ by the NA22-Collaboration and the data for $p\bar{p}$ collisions at $\sqrt{s} = 630\ GeV$ obtained by the UA1-Minimum-Bias-Collaboration. It was shown that the difference in the NA22- and the UA1-data cannot be accounted for by the different phase space cuts of the two experiments.

---

[7]A formalism which explicitly takes into account the effect of resonance decays on BEC is presented in Ref. [11].

As can be seen from Table 1, the fits to the data based on the model of a boost-invariant, longitudinally expanding source show that the fit parameters change significantly from the NA22- to the UA1-data. The transverse radius of the chaotic source, $R_{ch,\perp}$, increases by about 50% (from $\sim 0.7\ fm$ to $1\ fm$) and the correlation length in longitudinal direction $L_\eta$ is reduced by more than 50% (it drops from $0.9\ fm$ to $0.4\ fm$). On the other hand, $\tau_0$ is found to be almost unaffected (it changes from $1.9\ fm$ to $2\ fm$). The correlation length in transverse direction $L_\perp$ (which in the model considered here can be expressed in terms of $R_{ch,\perp}$ and the average transverse momentum of particles, cf. [2]) decreases from $\sim 0.8\ fm$ to $0.6\ fm$. If there is a coherent component, the transverse radius of the coherent source remains constant at a value of $R_{c,\perp} \sim 0.5\ fm$. Concerning the chaoticity parameter one can conclude that it either increases or remains approximately[8] constant.

The increase in the transverse radius $R_{ch,\perp}$ from the NA22- to the UA1-data may be related to the difference in the cross sections for the reactions $\pi^+ p/K^+ p$ at $\sqrt{s} = 22\ GeV$ and $p\bar{p}$ at $\sqrt{s} = 630\ GeV$, which is due to the dependence of the cross sections both on the reaction type and the collision energy. For the total cross sections, one has [12]

$$\frac{\sigma^{p\bar{p}}(22\ GeV)}{\sigma^{\pi^+ p}(22\ GeV)} \sim 1.7, \qquad \frac{\sigma^{p\bar{p}}(630\ GeV)}{\sigma^{p\bar{p}}(22\ GeV)} \sim 1.5 \qquad (11)$$

Thus, if one assumes that the transverse radius scales like the square root of the total cross section, one would expect an increase of $R_{ch,\perp}$ by about 60% which is compatible with our results (cf. Table 1). For a hydrodynamically expanding source the $s$-dependence of the longitudinal and transverse flow components may also contribute to the increase of the transverse radius. As was shown in Ref. [13], for reasonable values of the freeze-out temperature transverse flow can lead to a *decrease* of the transverse source radius extracted from BEC data (cf. Fig. 3 in [13]). At higher energies, the longitudinal expansion dominates and transverse flow becomes less important (cf. Ref. [14]). Therefore, the transverse source radius can be expected to increase with the CM energy.

---

[8]The data are also consistent with a slight decrease from $p \simeq 1$ at $\sqrt{s} = 22\ GeV$ to $p \simeq 0.9$ at $\sqrt{s} = 630\ GeV$.

The correlation lengths reflect dynamical properties of the hadronic medium rather than the geometry of the source and thus should not depend strongly on the reaction type. The data thus indicate a significant decrease of the correlation lengths $L_\eta$ and $L_\perp$ with increasing CM energy. This is the first time that information about the $s$-dependence of the correlation length has been extracted from BEC data [9].

On the theoretical side, there exist surprisingly few predictions concerning the dependence of BEC on the CM energy $\sqrt{s}$. A decrease of the width of the BEC function in longitudinal direction, $C_2(q_{||})$, with increasing $s$ was predicted by Bowler in the context of a string model[16]. A similar behaviour is also expected in a hydrodynamic scenario: for a longitudinally expanding source, the width of $C_2(q_{||})$ is approximately inversely proportional to the freeze-out proper time $\tau_f$ [17]. If the formation time $\tau_0$ of the thermalized matter is assumed to be $s$-independent, $\tau_f$ increases with the initial energy density which in turn increases with $s$. Consequently, the width of $C_2(q_{||})$ is expected to decrease with increasing CM energy.

Note that our results concerning the $s$-dependence of the space-time parameters of the expanding source are in qualitative agreement with the experimental observation [18, 19, 20, 21] that at fixed $\sqrt{s}$ the width of the BEC function decreases with incre-asing rapidity density $dn/dy$. Fluctuations in $dn/dy$ correspond to fluctuations in the hadronic energy $W$, i.e., the part of the CM energy that goes into the production of secondary particles. Since the average $dn/dy$ increases with $s$, the $dn/dy$-dependence of the BEC data suggests that the width of the correlation function decreases, which in the expanding source model is reflected in an increase of $R_{ch,\perp}$ and a decrease of $L_\eta$, $L_\perp$ with increasing CM energy. On the other hand, the data show that the "intercept" of the correlation function, as obtained from extrapolating a simple Gaussian form to $Q = 0$, considerably decreases with increasing $dn/dy$, which could be taken as an indication that the chaoticity decreases with $W$, and hence with $\sqrt{s}$. Such a behaviour (i.e. a significant decrease of $p$ with increasing $s$) is not consistent with the results of our present analysis. The resolution of this apparent discrepancy may be linked to the fact that the extraction of a reliable value for the chaoticity parameter requires knowledge of the behaviour of the BEC function near the origin, i.e., at small momentum differences. It would therefore be useful also from this point of view if the experimentalists could measure $\tilde{C}_2(Q^2)$ at fixed $dn/dy$ down to values of $Q \simeq 30\ MeV$. This kind of analysis would help to determine the dependence of the quantum statistical parameters on the *two* variables $dn/dy$ and $s$.

---

[9]In Ref. [15], an analysis of the dependence of the normalized factorial moments of the multiplicity distribution on the width of the rapidity interval for NA22- and for UA1-data led to the conclusion that the correlation length in rapidity increases with $s$. However, in [15] no distinction between long range and short range correlations was made. As was shown recently in [10], the long range fluctuations increase strongly with $s$, both in intensity and range. Thus there is no contradiction between the results of [15] and those of the present paper.

This work was supported by the Federal Minister of Research and Technology under contract 06MR731 and the Deutsche Forschungsgemeinschaft (DFG). B.R. Schlei acknowledges a DFG fellowship. We are indebted to B. Buschbeck (UA1 Collaboration), W. Kittel (NA22 Collaboration) and U. Ornik for many instructive discussions.

# References

[1] I.V. Andreev, M. Plümer, B.R. Schlei, R.M. Weiner, Phys. Lett. $\underline{B316}$ (1993) 583.

[2] I.V. Andreev, M. Plümer, B.R. Schlei, R.M. Weiner, Phys. Rev. $\underline{D49}$ (1994) 1217.

[3] B. Buschbeck, N. Neumeister, D. Weselka and P. Lipa, UA1-Minimum-Bias Collaboration, in *XXII International Symposium on Multiparticle Dynamics 1992*, ed. C. Pajaraes, World Scientific, Singapore, 1993, p. 246; N. Neumeister et al., UA1-Minimum-Bias-Collaboration, Z. Phys. $\underline{C60}$ (1993) 633.

[4] N. Agababyan et al., EHS/NA22 Collaboration, Z. Phys. $\underline{C59}$ (1993) 405.

[5] I.V. Andreev and R.M. Weiner, Phys. Lett. $\underline{253B}$ (1991) 416; Nucl. Phys. $\underline{A525}$ (1991) 527.

[6] I.V. Andreev, M. Plümer, and R.M. Weiner, Phys. Rev. Lett. $\underline{67}$ (1991) 3475; Int. J. Mod. Phys. $\underline{A8}$ (1993) 4577.

[7] I.V. Andreev, M. Plümer, B.R. Schlei and R.M. Weiner, Proceedings of the *XXIV Int. Symposium on Multiparticle Dynamics*, Vietri sul mare, Italy, Sept. 1994.

[8] B. Saleh, Photoelectron Statistics, Springer, Berlin, 1978.

[9] M. Gyulassy, S.K. Kauffmann, and L.W. Wilson, Phys. Rev. $\underline{C20}$ (1979) 2267.

[10] I.V. Andreev, M. Plümer, B.R. Schlei, R.M. Weiner, Phys. Lett. $\underline{B321}$ (1994) 277.

[11] J. Bolz, U. Ornik, M. Plümer, B.R. Schlei and R.M. Weiner, Phys. Rev. $\underline{D47}$ (1993) 3860.

[12] Particle Data Group, Phys. Lett. $\underline{B239}$ (1990).

[13] B.R. Schlei, U. Ornik, M. Plümer and R.M. Weiner, Phys. Lett. $\underline{B293}$ (1992) 275.

[14] U. Ornik, R.M. Weiner and G. Wilk, in *Proceedings of Quark Matter 93*, Nucl. Phys. $\underline{A}$, to appear.

[15] G.N. Fowler et al., J. Phys. $\underline{G16}$ (1988) 3127.

[16] M.G. Bowler, Z. Phys. $\underline{C29}$ (1985) 617.

[17] B. Lörstad and Yu.M. Sinyukov, Phys. Lett. $\underline{B265}$ (1991) 159; Yu.M. Sinyukov, Nucl. Phys. $\underline{A498}$ (1989) 151c.

[18] T. Akesson et al., Phys. Lett. $\underline{B129}$ (1983) 269; Phys.Lett. $\underline{B155}$ (1985) 128; Phys.Lett. $\underline{B187}$ (1987) 420; A. Breakstone et al., Phys. Lett. $\underline{B162}$ (1985) 400; Z. Phys. $\underline{C33}$ (1987) 333.

[19] C.-E. Wulz, in *Multiparticle Production (Proceedings of the Perugia Workshop)*, eds. R. Hwa, G. Pancheri and Y. Srivastava, World Scientific, Singapore, 1989, p. 41; C. Albajar et al., Phys. Lett. $\underline{B226}$ (1989) 410.

[20] T. Alexopoulos et al., Phys. Rev. $\underline{D48}$ (1993) 1931.

[21] N.M. Agababyan et al., EHS/NA22 Collaboration, Z. Phys. $\underline{C59}$ (1993) 195.

# Figure Captions

**Fig. 1** Two particle correlation function $\tilde{C}_2(Q^2)$, for (a) $K^+p/\pi^+p$ collisions at $\sqrt{s} = 22\ GeV$ and (b) $p\bar{p}$ collisions at $\sqrt{s} = 630\ GeV$. The fits to the UA1-data [3] and the NA22-data [4] were obtained for a purely chaotic, boost-invariant longitudinally expanding source.

# Table Captions

**Table 1** Fit parameter values and $\chi^2/d.o.f.$ for the expanding source model.

| Expanding source | | | | | |
|---|---|---|---|---|---|
| $\sqrt{s}$ [GeV] | $p$ | $\tau_0$ [fm] | $R_{ch,\perp}$ [fm] | $L_\eta$ [fm] | $\chi^2/d.o.f.$ |
| 630 | 1.00 | $2.12^{+0.07}_{-0.08}$ | $1.03^{+0.01}_{-0.03}$ | $0.42^{+0.01}_{-0.02}$ | 49.77/43 |
| 22 | 1.00 | $1.88^{+0.07}_{-0.12}$ | $0.63^{+0.04}_{-0.03}$ | $0.88^{+0.03}_{-0.05}$ | 36.97/36 |
| 22 | 0.60 | $1.92^{+0.13}_{-0.13}$ | $0.73^{+0.03}_{-0.04}$ | $0.90^{+0.05}_{-0.06}$ | 31.76/36 |

# Table 1

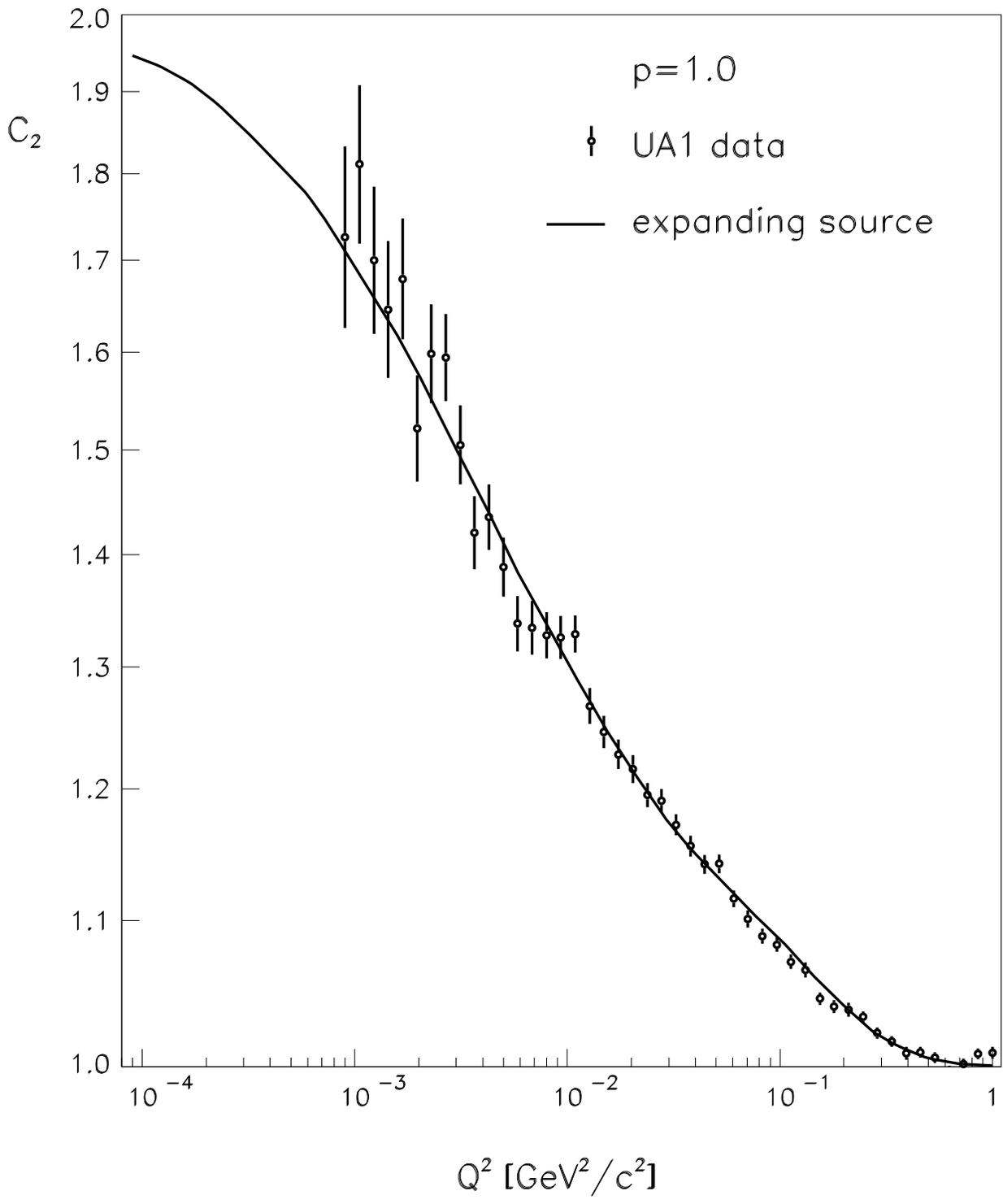

Figure 1a

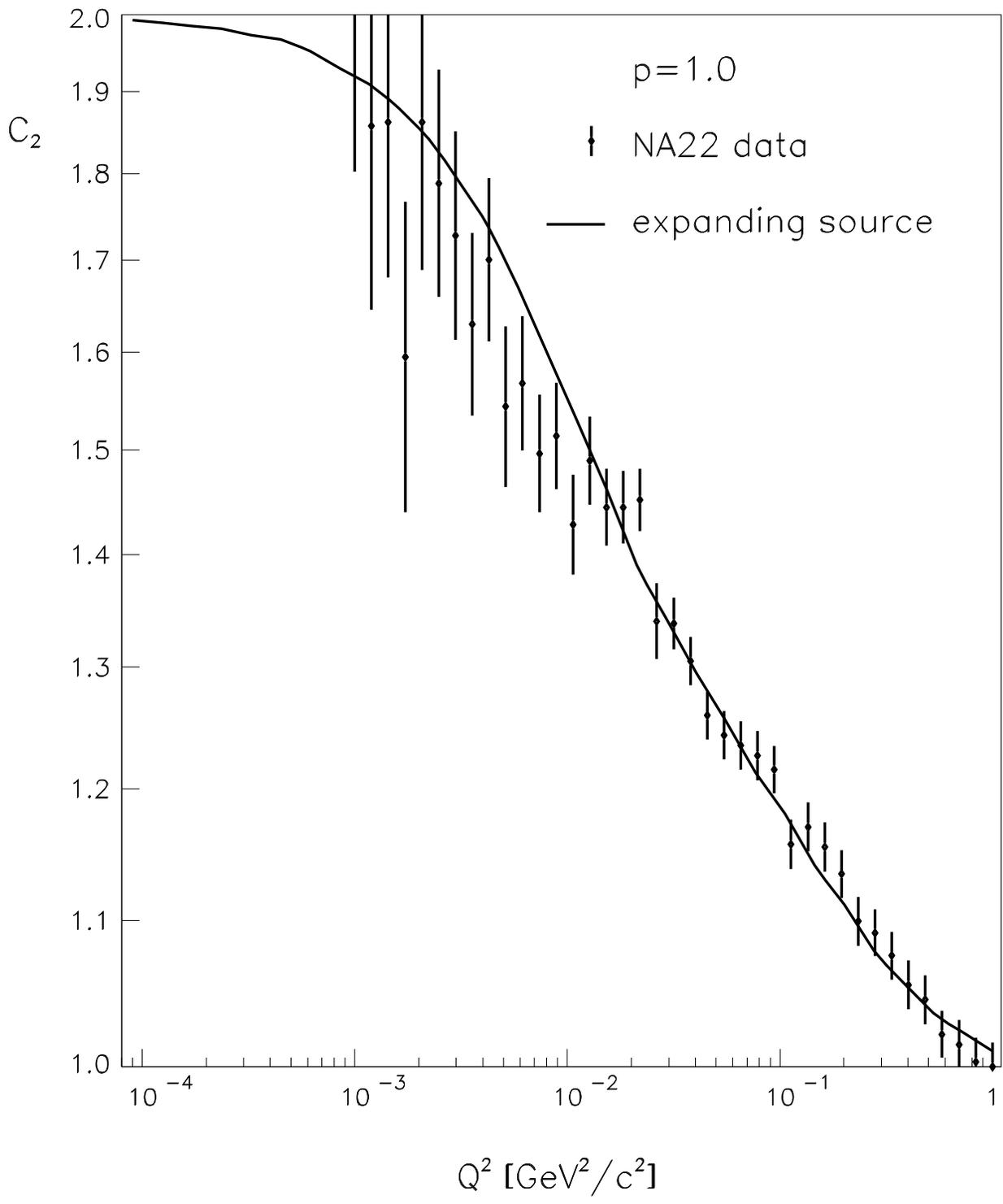

Figure 1b